\documentclass[]{spie}  

\newcommand{\hr}{HR~4796A}
\newcommand{\farcsec}{\hbox{$.\!\!^{\prime\prime}$}}

\newcommand{\iband}{$i'$-band}

\newcommand{\rband}{$r'$-band}

\newcommand{\adi}{KLIP-ADI}

\usepackage{amsmath,amsfonts,amssymb}
\usepackage{graphicx}
\usepackage[colorlinks=true, allcolors=blue]{hyperref}
\usepackage[symbol]{footmisc}

\title{WindsoCC: reconstructing the wind-driven halo in MagAO-X images using wavefront sensor telemetry\footnote[2]{\hspace{3pt} This paper includes data gathered with the 6.5 meter Magellan Telescopes located at Las Campanas Observatory, Chile.}}

\author[a,b,*]{Jay K. Kueny}
\author[a]{Jared R. Males}
\author[e]{Alycia J. Weinberger}
\author[a]{Laird M. Close}
\author[d]{Joseph D. Long}
\author[a,b]{Joshua Liberman}
\author[f,a]{Sebastiaan Haffert}
\author[a,b,c]{Eden McEwen}
\author[a]{Maggie Y. Kautz}
\author[g,a,b]{Olivier Guyon}
\author[h]{Logan Pearce}
\author[a,c]{Parker T. Johnson}
\author[a,b]{Katie Twitchell}
\author[a,c]{Jialin Li}
\author[i]{Alex Hedglen}
\author[j]{Avalon Gower}
\author[a]{Warren Foster}
\author[k]{Jhen Lumbres}
\author[l]{Lauren Schatz}

\affil[a]{Steward Observatory, University of Arizona, 933 N Cherry Ave, Tucson, AZ 85721, USA}
\affil[b]{James C. Wyant College of Optical Sciences, University of Arizona, 1630 E. University Blvd., Tucson, AZ 85721, USA}
\affil[c]{National Science Foundation Graduate Research Fellow}
\affil[d]{Center for Computational Astrophysics, Flatiron Institute, 162 5th Ave, New York, NY}
\affil[e]{Earth and Planets Laboratory, Carnegie Institution for Science, 5241 Broad Branch Road NW, Washington, DC 20015-1305}
\affil[f]{Leiden Observatory, Leiden University, PO Box 9513, 2300 RA Leiden, The Netherlands}
\affil[g]{Subaru Telescope, National Observatory of Japan, Hilo, HI}
\affil[h]{Department of Astronomy, University of Michigan, Ann Arbor, MI}
\affil[i]{Northrop Grumman Corporation, 600 South Hicks Rd, Rolling Meadows, IL}
\affil[j]{Draper Laboratory, 555 Technology Square, Cambridge, MA}
\affil[k]{Northrop Grumman, Pasadena, CA}
\affil[l]{Starfire Optical Range, Kirtland Air Force Base, Albuquerque, NM}

\authorinfo{Send correspondence to J. K. Kueny \\ J. K. Kueny: E-mail: jkueny@arizona.edu}

\begin{document} 
\maketitle

\begin{abstract}
The wind-driven halo (WDH) is a persistent, low spatial frequency noise artifact that arises due to the servo-lag error inherent to all adaptive optics (AO) instruments. Spatial filtering may be employed to overcome this artifact, however, filtering out the WDH while simultaneously preserving signal from an extended astrophysical object of interest is exceptionally challenging. Additionally, since the WDH changes in intensity and position angle through an observation, data-driven algorithms (e.g., KLIP) that are commonly used to subtract the starlight need to be overly-aggressive to remove both the static and dynamic noise components. Since wavefront sensors (WFSs) continuously track the closed-loop residual wavefront error, WFS telemetry presents the ideal resource for combating this type of noise artifact through postprocessing. Using archival WFS telemetry from MagAO-X, which is the ``extreme" AO instrument for the 6.5-meter Magellan-Clay telescope, we demonstrate a novel workflow for WDH reconstruction and removal in individual coronagraphic science images. MagAO-X is equipped with a pyramid WFS capable of recording wavefront telemetry at a high-cadence which is saved during data acquisition. Given this, we detail how our WFS data processing pipeline, \texttt{windsoCC}, cross-correlates the recorded closed-loop wavefront to measure the wind vectors of several turbulent layers of the atmosphere above Las Campanas Observatory. We then make use of the wind parameters learned through \texttt{windsoCC} to reconstruct the WDH footprint by leveraging a parametric model. Notably, we demonstrate a dramatic improvement in object recovery using on-sky MagAO-X images of the disk around HR~4796A at visible wavelengths. 
\end{abstract}


\section{INTRODUCTION}
\label{sec:intro}  

To aid in the study of circumstellar objects at high-contrast, almost all of the largest ground-based optical telescopes are assisted with adaptive optics (AO). AO facilitates diffraction-limited imaging from the ground, and so telescope aperture sizes in the 6 to 10 meter class can be particularly effective at directly imaging the regions around nearby stars that contain, e.g., giant planets and Kuiper belt analogs. Maximizing the achievable contrast at small inner-working angles (IWAs) is of great interest for astronomers looking to characterize the innermost regions of nearby circumstellar environments (e.g., Ref.~\citenum{balsalobre-ruzaTwoInnerDust2026}).

The wind-driven halo (WDH) is a pervasive source of starlight leakage within the AO control region and can severely limit contrast. It is a dynamic and high-contrast noise artifact \cite{cantalloubeWinddrivenHaloHighcontrast2020}, which makes it difficult to remove through post-processing. The WDH is often the most limiting source of noise in high-contrast images, especially images of extended objects like circumstellar disks. For circumstellar disks, high-pass filtering for WDH mitigation is ill-advised because it is easy to also filter out smooth, diffuse disk features.

For a given observation, attaining a full understanding of how the WDH affects the science image data requires knowing how the turbulence behaved during data acquisition. One way to profile the turbulence during an observation is using wavefront sensor (WFS) telemetry. Under Taylor's ``frozen-flow" hypothesis \cite{taylorSpectrumTurbulence1938}, layers of wind within the telescope's line-of-sight will appear as 2D phase screens undergoing translation across the pupil. In this scenario, and specifically for a pyramid WFS, applying the cross-correlation (CC) operation to the telescope pupil images in the closed-loop WFS data can extract wind vectors for each turbulent layer. Namely, by applying the CC of a given WFS pupil image with itself across a range of temporal delays, turbulent layers would manifest as moving peaks within the resulting CC response maps (e.g., Ref.~\citenum{schockMethodQuantitativeInvestigation2000}). These moving peaks encode the turbulent layer's speed and direction. This information can then be matched to the science image data taken at the same time to inform the orientation of the WDH in the science images.

Portions of this work, including the detailed WDH forward modeling and pipeline architecture, have been submitted to the Journal of Astronomical Telescopes, Instruments, and Systems (JATIS) for peer review.

\section{Observations}
\label{sec:observations}

\begin{table*}[!ht]
\centering
\small
\caption{MagAO-X Observation Log of \hr{}}
\begin{tabular}{lcclcccccc}
\hline
Date        & Time         & Seeing$^a$  & Band  & Air Mass & $t_{\text{exp}}$ & $N_{\text{exp}}$ & $N_{\text{coadds}}$ & $\theta$ &  \\
            & (UT)         & $(")$       &       &          & (s)              &                  &                     &  $(^{\circ})$ \\ \hline
2023 Mar 10 & 05:49/07:36 & 0.6 &  $i'/z'$  & 1.0/1.1  & 1                    & 5047    &     84         & -21.5/65              &            \\
2023 Mar 13 & 04:46/07:15 & 0.6 & $r'/g'$  & 1.1/1.0  & 0.23                 & 35355   &   122               & -58.5/61.5             &             \\

\hline
            &              &            &       &          &                      &                     &                  &                   &                  
\end{tabular}

\footnotesize
\raggedright
\textbf{Notes.} Time: start/end observation; Air Mass: start/end air mass; $t_{\text{exp}}$: single frame exposure time; $N_{\text{exps}}$: number of raw science images; $N_{\text{coadds}}$: number of image cutouts used for PSF subtraction;  $\theta$: start/end parallactic angle. The total integration time for a given dataset is found by multiplying $t_{\text{exp}}$ and $N_{\text{exp}}$. \\
\vspace{0.5mm}
$^a$ median values in arcseconds.
\label{tab:obslog}
\end{table*}

We observed \hr{} on 2 separate nights using the 6.5 m Magellan-Clay telescope at Las Campanas Observatory (LCO) during the 2023A semester. MagAO-X operates with the pupil fixed with respect to the sky to facilitate ADI; see Table \ref{tab:obslog}). Additionally, we made use of MagAO-X's small Lyot coronagraph (focal plane mask radius of $\lesssim3 \lambda/D$) for all observations. The instrument has a plate scale of 0\farcsec0059 \cite{longAstrometricCalibrationMagAOX2025} and a field-of-view of $6\farcsec0 \times 6\farcsec0$.

MagAO-X has a suite of 3 deformable mirrors (DM) with the 2k-actuator DM operating in the high-order AO loop acting as the workhorse of the instrument. Using this DM, we can generate four additional artificial DM speckles (see Ref.~\citenum{mcewenOnskyRealtimeOptical2024}) by setting a sinusoidal pattern on the face sheet that are useful for, e.g., astrometry, photometry, and frame registration. These artificial speckles offer a convenient way to measure the Strehl ratio of the PSF during coronagraphic observations and so we used these image features to exclude frames of poor quality (explained in Section \ref{sec:forward-modeling}). For our \hr{} observations, we generated speckles at a distance of $15 \lambda/D$ from the star.

\section{\texttt{windsoCC} pipeline description}
\label{sec:pipeline}

\begin{figure}[ht]
\includegraphics[width=\linewidth]{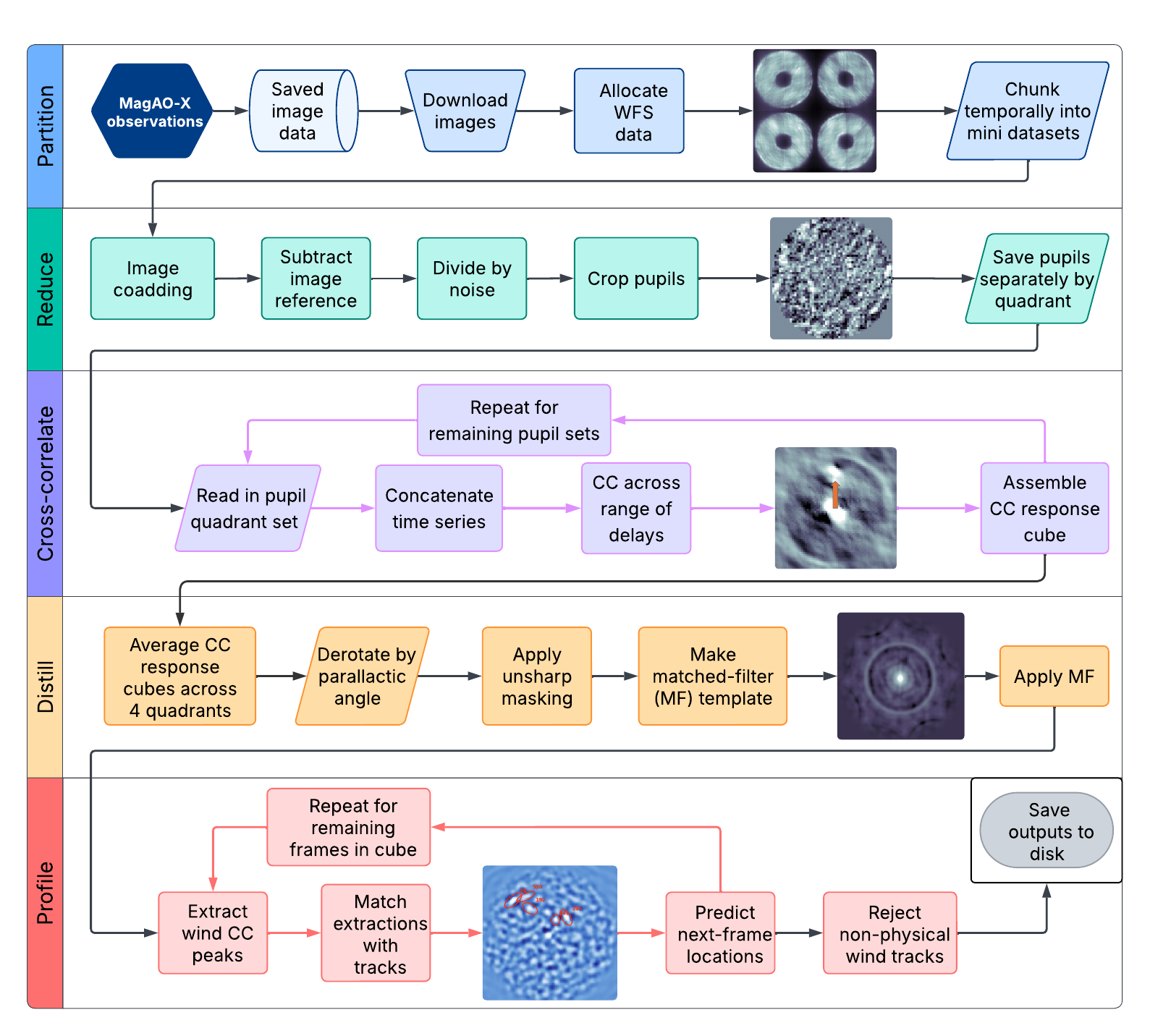}
\caption
{Flowchart illustrating the various operations performed in the \texttt{windsocc} pipeline. This workflow is explained in detail in Section \ref{sec:pipeline}. We inset small thumbnail images illustrating data products at the intermediate steps. We use standard ANSI flowchart shapes to denote processing steps, manual operations, data products, etc. The final outputs include a human-readable text file that contains the information on each detected wind layer and the MF response cubes.}
\label{fig:flowchart}
\end{figure}

In order to gauge how different layers of atmospheric turbulence affect MagAO-X's performance, we implemented a pyramid WFS \cite{ragazzoniPYRAMIDWFS1996} image processing pipeline, \texttt{windsoCC}, based on CC. The basic premise is that under the assumption of frozen-flow, the telescope pupil in the WFS images contain translating patterns due to layers of wind. One can then apply the CC on a given pupil image with itself at a later time to create a CC response map that contains distinct peaks that are at some distance from the origin. Performing the CC iteratively across a range of delays should then produce response maps that show peaks moving radially outward from the image center with increasing temporal delay. Then, these peaks can be analyzed to measure a given wind layer's speed and direction. Note that because we are using closed-loop WFS telemetry, we can leverage the \texttt{windsoCC} output to inform our WDH model parameter priors when reconstructing the WDHs. Our pipeline follows a number of discrete stages that we illustrate using a flowchart in Figure \ref{fig:flowchart}. A high-level overview of these various pipeline stages continues below.

The MagAO-X pyramid WFS records telemetry at a cadence of up to 3.6kHz \cite{malesMagAOXCommissioningResults2024}, producing a large number of image data FITS files over the course of a science observation. As a first step, \texttt{windsoCC} partitions the total WFS image dataset (typically encompassing hours of integration time) into mini datasets spanning 12 seconds of integration time. For each of these mini datasets, the pipeline normalizes the images by subtracting off a reference and dividing by an estimate of the noise; the former is made by the averaging and the latter is made by computing the standard deviation for each pixel using all of the images in the mini dataset. \texttt{windsoCC} then cuts out each of the four pupil images and organizes them in folders corresponding to each quadrant.

The next step is applying the CC to each pupil quadrant folder separately. To maximize the signal-to-noise of the CC peaks due to turbulent wind, \texttt{windsoCC} uses a segmented, overlapping CC method, conceptually analogous to the variance-reduction techniques used in Welch's method for power spectral density estimation \cite{welchUseFastFourier1967}. Within each mini dataset, a maximum correlation time delay ($\Delta t_{max}$) of 2 seconds was used for this dataset to capture the slow-moving ground layers. The reduced frames in this block are subdivided into a series of overlapping segments of duration $\Delta t_{max}$; our CC method uses a 50\% overlap between adjacent segments. For example, the first segment comprised the frames spanning $t = 0$ to $2$ seconds, the second segment spanned $t = 1$ to $3$ seconds, the third $t = 2$ to $4$ seconds, and so forth. Then, the CC is performed over a configurable range of frame delays. The CC was computed independently for each 2-second segment, generating a localized response cube. Finally, the response cubes for all overlapping segments within the 12-second block were temporally averaged.

After the WFS data have been CCed, the next step is to combine the four quadrants associated with each image timestamp into a single, CC response map cube. This ``distill" stage then yields a single time series of length $\Delta t_{max}$ of CC response maps. To further increase the signal-to-noise ratio of the CC peaks due to the wind, \texttt{windsoCC} applies unsharp masking and matched filtering (MF). The matched-filter template is the first slice in the CC response cube.

Finally, the last stage of the pipeline involves detecting and tracking the peaks in the CC response maps that are due to the layers of turbulence. To do this, \texttt{windsoCC} defines an annular ``tripwire'' region around the central autocorrelation peak whose width is configurable where \texttt{SEP} will search for peaks and ignore the rest of the image. The concept of the tripwire search region was motivated by the idea that all peaks due to a wind layer will traverse this region going linearly, radially-outward, and at a constant velocity. Peak detections that are due to noise will not follow these 3 rules and thus become easy to dynamically reject. For these data, we set the tripwire region to span the region between 12 and 24 pixels from the autocorrelation peak (see Figure \ref{fig:flowchart}, ``Distill" lane for an inset response map with the central autocorrelation peak).

The final \texttt{windsoCC} measurement products summarize each accepted wind track with a small set of physically interpretable parameters. For every matched-filter response cube, the pipeline reports the wind layer track identifier, wind speed, propagation direction, and the peak's average area in pixels. The resulting files form a time-resolved lookup table of exceptionally-turbulent wind layers during the observation.

\section{Retrieved wind vectors}
\label{sec:retrieved-wind-vectors}

\begin{figure}
\begin{center}
\includegraphics[width=0.65\linewidth]{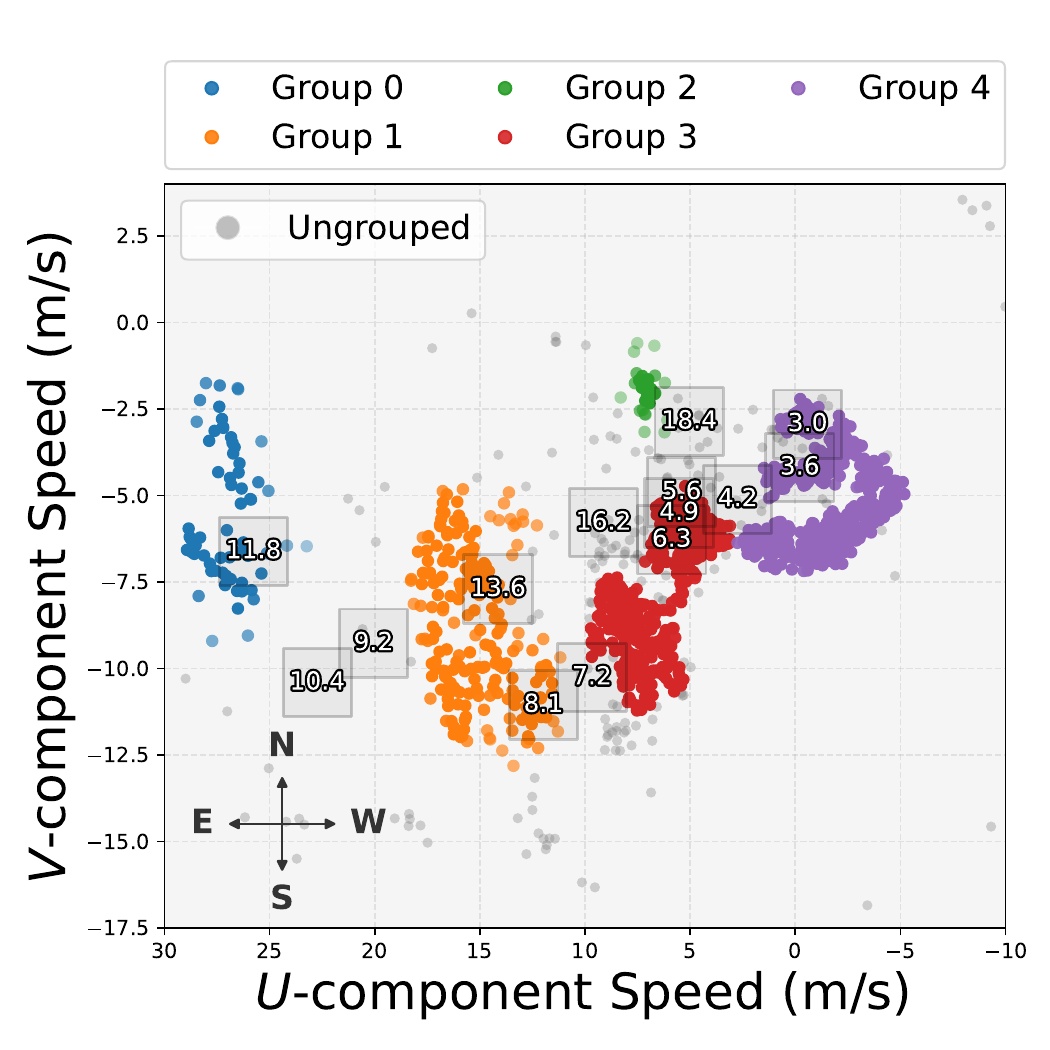}
\end{center}
\caption{Clustering analysis on wind layer velocities and directions extracted by the \texttt{windsocc} pipeline using the dataset collected on UT 2023 Mar 12. The vertical and horizontal axes show the V-component ($v\sin\theta$) and U-component ($v\cos\theta$) of the wind direction vectors respectively. We used the Hierarchical Density-Based Spatial Clustering of Applications with Noise (HBDSCAN) algorithm to facilitate cluster identification in this feature space. We assign colors to the identified groups according to the legend on top of the plots and the relative opacity of the assigned colors represents the probability of individual points belonging to that colored cluster. We illustrate points identified as noise by HBDSCAN as light gray dots. For reference, we plot the altitudes in kilometers of wind properties using historical wind data from the ERA5 dataset.}
\label{fig:clusters-wind}
\end{figure}

We utilized the Hierarchical Density-Based Spatial Clustering of Applications with Noise (HDBSCAN) algorithm on the \texttt{windsoCC} output to classify discrete layers of atmospheric turbulence over the course of our observations. We show the results for the dataset collected on UT Mar 12 2023 in Figure \ref{fig:clusters-wind}. Cross-referencing our empirical measurements with historical weather data from the Copernicus ERA5 reanalysis dataset \cite{hersbachERA5GlobalReanalysis2020} confirmed that many of the extracted wind vectors aligned with the multi-layer atmospheric profile above Las Campanas Observatory at the time of observation. We show the historical wind properties at various altitudes as framed numbers overlaid on the scatter plot area in Figure \ref{fig:clusters-wind}.

\section{Forward modeling the wind-driven halo}
\label{sec:forward-modeling}

\begin{figure}
\centering
\includegraphics[width=0.8\linewidth]{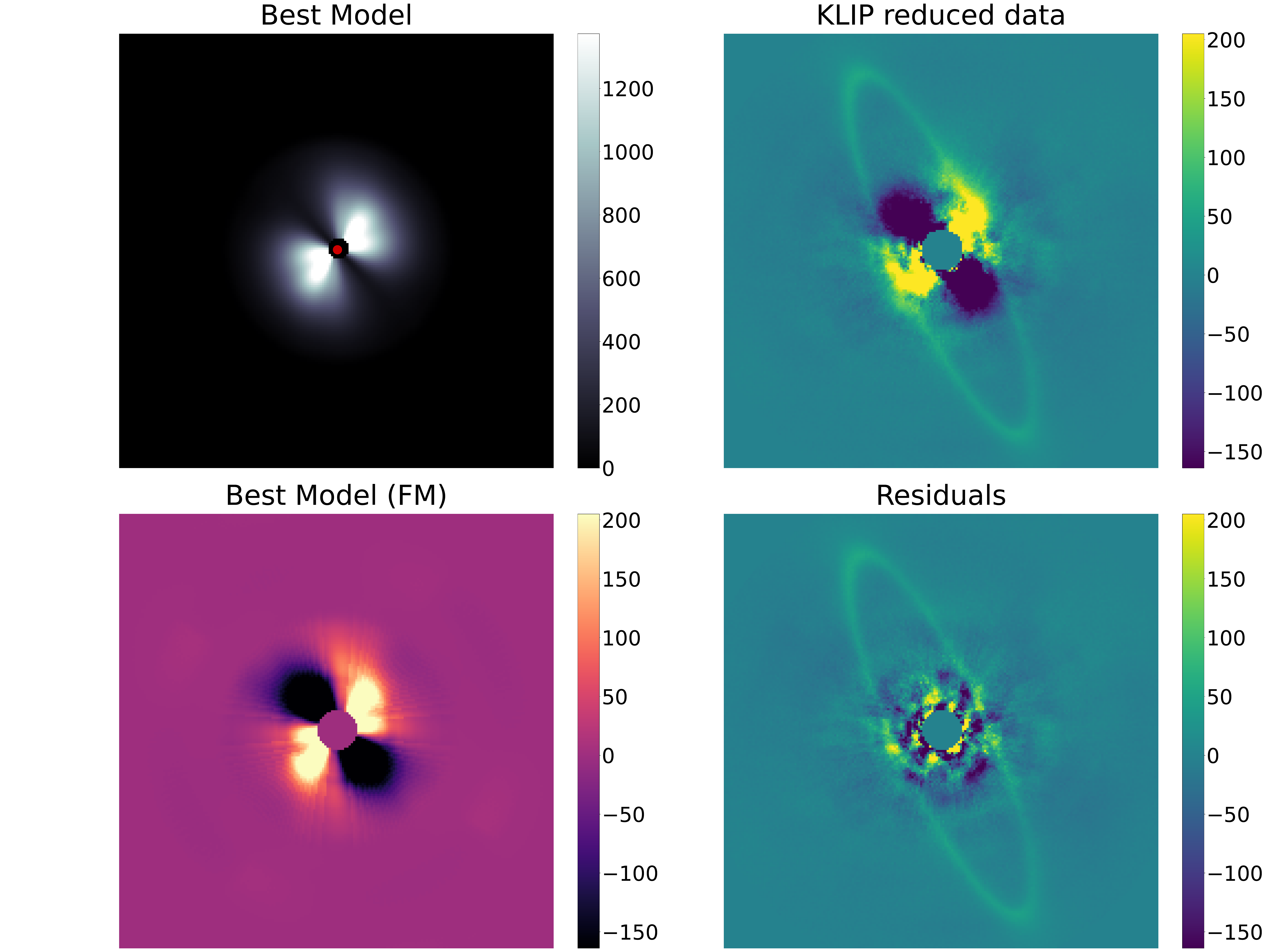}
\caption
{WDH forward modeling results using the KLIP-reduced image of \hr{} from the observations collected on UT 2023 Mar 12. Top left: The best-fitting model consisting of a sum of 4 individual WDH models. Top right: KLIP-reduced image made using 1 KLIP mode. Bottom left: The best WDH model after forward-modeling through the KLIP algorithm. Bottom right: Residuals image made from subtract the best forward model from the KLIP-reduced image.}
\label{fig:windfm-r}
\end{figure}

Because the WDH is a dynamic, high-contrast feature composed of multiple turbulent layers, the WDH footprint in individual MagAO-X science images is weak. However, the WDH artifact in the final reduced image is very distinct. Therefore, we opted to reduce our \hr{} images using conservative Karhunen-Loéve Image Projection (KLIP) parameters and then explicitly forward-model a parametric WDH artifact utilizing a Markov Chain Monte-Carlo (MCMC) framework. We used the \textsc{DiskFM} tool \cite{mazoyerDiskFMForwardModeling2020} (part of the \texttt{pyKLIP} package \cite{wangPyKLIPPSFSubtraction2015}) to handle the forward modeling through the KLIP algorithm.

We implemented a 2D analytical model to mimic the morphology of the WDH, incorporating a Gaussian envelope along the wind direction, a pinch parameter controlling the opening angle, and a parameter to govern radial flaring:

\begin{equation}
    I(x,y) = N_0 \exp{\left[-\frac{1}{2}\left(\frac{r}{p_0^{\gamma}} \right)^2\right]}
    \begin{cases}
        \exp{-\frac{1}{2}\left(\frac{x}{\sigma_+} \right)^2 }; & x > 0 \\
        \exp{-\frac{1}{2}\left(\frac{x}{\sigma_-} \right)^2 }; & x < 0,
    \end{cases}  
\end{equation}

where $r^2 = x^2 + y^2$, $p_0$ is the pinch parameter which controls the opening angle of the lobes, $\sigma_+$ and $\sigma_-$ control the brightness of individual lobes to fit asymmetries, $\gamma$ controls the radial flaring of the lobes, and $N_0$ is a scale factor. We rotated the coordinate system to align with the orientation vectors identified by \texttt{windsoCC} using another configurable parameter $\theta_{PA}$ before generating the model:

\begin{equation}
    \begin{split}
        x_{PA} = \hat{y}\sin\theta_{PA} - \hat{x}\cos{\theta_{PA}}, \\ y_{PA} = \hat{y}\cos\theta_{PA} + \hat{x}\sin\theta_{PA}.
    \end{split} 
\end{equation}

Finally, we generate our models in an annulus with the outer boundary extending just beyond the AO control region which has a diameter of $24 \lambda/D$ (e.g., for $r'$-band, we set the outer boundary to 500 mas). The inner boundary was set to 120 mas to exclude the innermost, high speckle noise region in the image.

\begin{figure}[ht!]
\begin{center}
\begin{tabular}{c}
\includegraphics[width=0.8\linewidth]{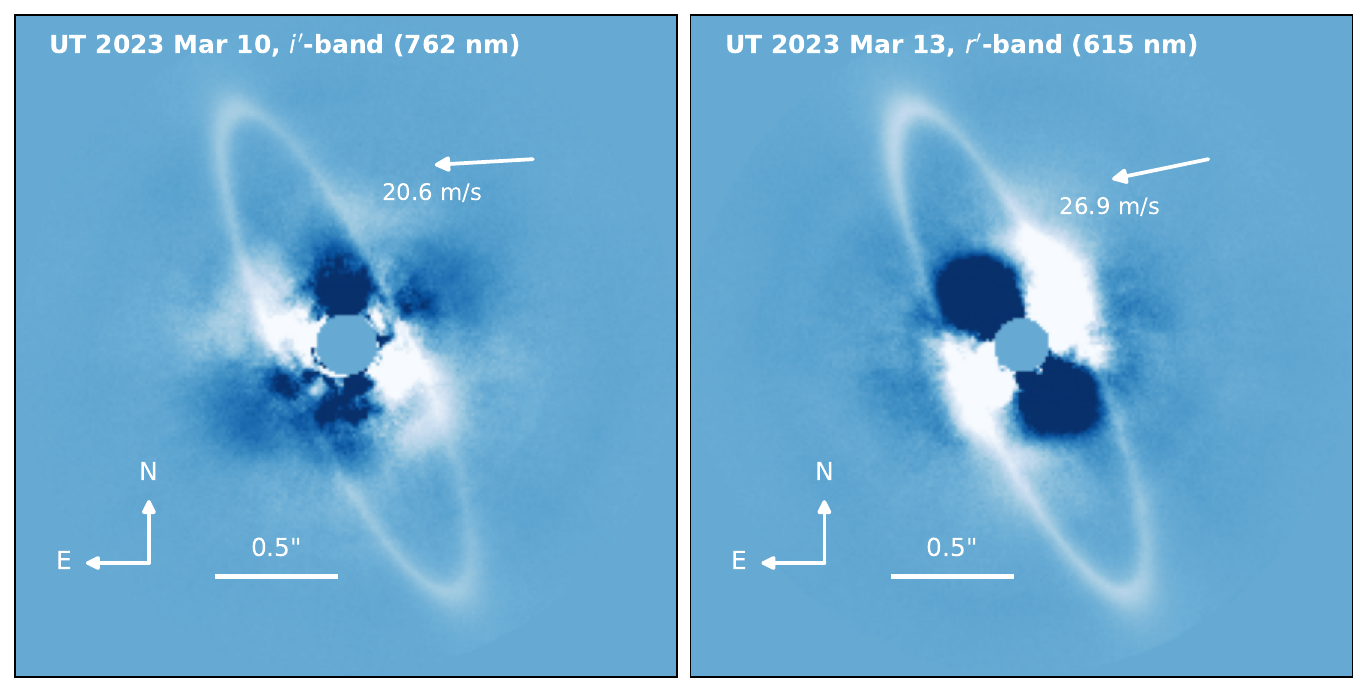} \\
\includegraphics[width=0.8\linewidth]{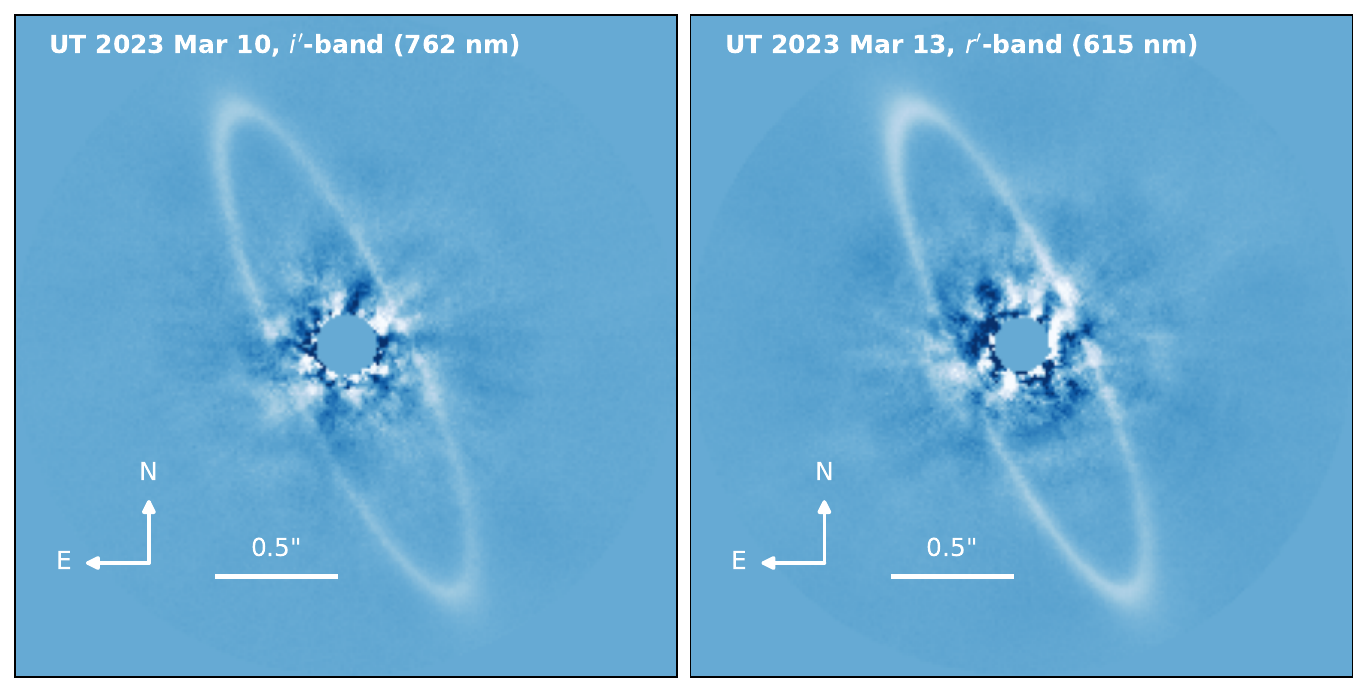}
\end{tabular}
\end{center}
\caption
{\textbf{Top row:} Example of the effect of the WDH on MagAO-X images reduced using \adi{}. The top left panel is the reduced image of \hr{} at \iband{} using data collected on UT 2023 Mar 10. The top right panel shows the reduced \rband{} image using data collected on UT 2023 Mar 13. We subtracted the PSF from both images using 1 KLIP mode. For both images, overlay the wind vector corresponding to the jet stream as the white arrow. \textbf{Bottom row:} Reduced images of \hr{} after removing the WDH from individual science frames and then subtracting the PSF using a single KLIP mode.}
\label{fig:wdhsub}
\end{figure}

To estimate how the WDH survives post-processing, we integrated our parametric models into \textsc{DiskFM} while utilizing the wind vectors extracted from our WFS telemetry to inform the parameter priors (such as the number of simultaneous WDH models and their expected position angles). For our UT 2023 Mar 10 and Mar 13 datasets, we fit combinations of 4 and 5 separate WDH models, respectively, to match the distinct clusters retrieved by our pipeline. We determined that the walkers had converged after the number of iterations had exceeded 50 times the autocorrelation length of the walker chains, which was achieved after testing 213k and 746k models for the UT 2023 Mar 10 and UT 2023 Mar 13 datasets, respectively.

\section{WDH removal and results}
\label{sec:results}

Following convergence of the MCMC walkers, we extracted the individual model components that constituted the total best-fit forward model. These best-fit components were compiled into a modal basis alongside a median-collapsed PSF estimate. Utilizing a non-negative least-squares fitting routine, we evaluated this basis against each individual coronagraphic frame to iteratively scale and subtract the WDH artifacts.

Our results demonstrated dramatic improvements. By scrubbing the WDH components directly from the individual science frames before final processing, we retrieved clear images of the \hr{} disk at $r'$-band (615 nm) and $i'$-band (762 nm) without relying on overly aggressive spatial filtering or PSF subtraction. This methodology allows extended object signals to remain preserved in the AO control region, establishing closed-loop WFS telemetry as an incredibly valuable asset for high-contrast imaging.

\section{Summary}
\label{sec:summary}

In summary, we built the automated \texttt{windsoCC} pipeline leveraging the CC operation on closed-loop pyramid WFS telemetry to empirically map the atmospheric wind vectors throughout MagAO-X operations. By bridging this telemetry with a custom forward-modeling framework (using \textsc{DiskFM}), we successfully parameterized and removed complex, multi-layered WDH artifacts from MagAO-X focal plane images. Ultimately, this allowed for cleaner recovery of the \hr{} debris disk and highlights the importance of retaining telemetry for advanced post-processing techniques.

\section{Acknowledgements}
\label{sec:acknowledgements}

J.K.K., J.R.M., Y.D.L., and A.J.W. acknowledge support from the NSF, grant no. AST-2307613. J.D.L. was supported by the Flatiron Software Research Fellowship at the Flatiron Institute, a division of the Simons Foundation.

This research contains modified Copernicus Climate Change Service information [2026]. Neither the European Commission nor ECMWF is responsible for any use that may be made of the Copernicus information or data it contains.

We are very grateful for support from the NSF MRI Award No. 1625441. The Phase II upgrade program is made possible by the generous support of the Heising-Simons Foundation. MagAO-X uses the CACAO software package, which is supported by NSF Award No. 2410616.

J.K.K. and J.R.M. acknowledge support from the NSF ATI Grant No. 2308351 for PSF reconstruction.

This material is based upon High Performance Computing (HPC) resources supported by the University of Arizona TRIF, UITS, and Research, Innovation, and Impact (RII) and maintained by the UArizona Research Technologies department.

\bibliography{report} 
\bibliographystyle{spiebib} 

\end{document}